\documentclass[conference]{IEEEtran}
\usepackage{cite}
  \usepackage[pdftex]{graphicx}
\usepackage[caption=false,font=footnotesize]{subfig}
\usepackage{txfonts}  
\usepackage{stfloats} \fnbelowfloat
\usepackage{url}
\usepackage{tikz}
\usetikzlibrary{patterns}
\begin{document}
\title{Power Supply Options for a Naval Railgun}
\author{\IEEEauthorblockN{S.\,Hundertmark, O.\,Liebfried}
\IEEEauthorblockA{French-German Research Institute,
Saint Louis, France\\
}
}

\maketitle
\begin{abstract}
Large railguns require powerful power supply units. At the French-German Research
Institute of Saint-Louis (ISL) most experimental railguns are driven by power supply units based on
capacitors. Recent investigations at ISL explore the possibility to use coil based
systems to increase the energy density of the power supply. In this study an electrical
circuit simulation is used to investigate the difference for railgun operation in between a 
capacitor and a coil based power supply with respect to current amplitude behavior and 
projectile velocity. For this a scenario of a 25\,MJ muzzle energy railgun is simulated
with two different power
supply options, replacing capacitors by coils and using a range of circuit resistances.
The resistance determines to a large part the losses of the system and defines
therefore the efficiency of the launch and the size of the power supply. The interpretation of 
the results of the performed simulations leads to the conclusion that the capacitor based system 
"naturally" produces a favorable current pulse trace with respect to launching a mechanical delicate payload.
Further simulations show that the disadvantage of the inductor based supply can be mitigated by
increasing the power supply unit subdivision into smaller units.
\end{abstract}
\section{Introduction}
Railguns are able to convert electrical energy into kinetic energy at the gigawatt power
level \cite{launch_effi}. Recent progress in railgun research allows to reach muzzle 
energies in excess of those being achieved by currently installed naval deck guns. Since
several years, efforts in the US to further develop such a gun and to mature the technology to
a useable, naval system have made tremendous progress. In \cite{mcnab_dev} and
\cite{mcnab_nav} parameters of the investigated gun system and possible application
scenarios are described. In the wake of this progress, the French-German Research Institute (ISL) 
started to investigate a shipboard long-range artillery scenario. Within this research
effort, it was demonstrated in the laboratory that velocities above 3\,km/s are achievable
\cite{launch_effi}. Further on, a preliminary launch package design was developed and
tested to show that hypervelocity projectiles can be launched using railguns
\cite{launch_pack}. In the long-range artillery scenario, the projectile is launched under
a steep firing angle to reach the distant target on a ballistic trajectory. Due to a muzzle velocity above
2000\,m/s such a projectile can cover target distances far above 100\,km. In
\cite{shipboard} a 25\,MJ muzzle energy railgun was investigated. Using a 6.4\,m long
barrel and a 4\,MA current allows to launch a mass of 8\,kg to 2500\,m/s. Not answered was the
question how a pulsed power supply  (PPS) could look like for the envisioned system. Two 
possible choices are a capacitor or an inductor based system. The incentive to use an
inductor based system is the higher energy density, possibly resulting in a
reduced footprint of such a PPS. In experiments reported by \cite{oli_1} it was shown 
than an energy density gain of more than 10 can be realized with coils compared to high-end capacitors.
Using an electrical simulation code, the effects of the two different PPS systems on the
railgun launch performance are investigated.
\section{General Considerations for the PPS}
A railgun uses electric current to drive the armature through the acceleration volume. A
military payload might be sensitive to changes in acceleration and therefore it is of
importance to aim for a constant current amplitude during the launch. The launch of a
heavy projectile requires the conversion of electrical energy of the order of 100\,MJ or more. Such an
amount of energy is usually not stored in a single cell, instead the PPS is subdivided
into several smaller, identical units. To generate a close to constant current amplitude
shape these units are triggered following a time sequence. One possibility to trigger the
release of a unit is to monitor the current amplitude and activate a subsequent unit
whenever the current falls below a certain value. This approach requires a measurement of
the total current and its interpretation. Usually applied in railgun experiments is another 
possibility: magnetic field sensors are placed along the barrel length and the passage of the 
armature is used to trigger one of the PPS units. At the same time the signal from the magnetic 
field sensors can be used to calculate the velocity of the armature. To obtain the velocity 
information during launch is essential, as it allows to reduce muzzle velocity dispersion
by slightly modifying the trigger instants during launch \cite{thorbjoern_1}. A natural way to place
these sensors is to do so by a fixed distance in between two sensors. Why this equidistant
spacing makes sense can be understood by investigating the required electrical power and
energy for the acceleration. When the armature traverses a certain length of an ideal railgun
(meaning no losses), the PPS has to supply energy to build up the magnetic field ($\Delta
E_{mag}$) behind the armature and to increase the kinetic energy ($\Delta E_{kin}$) of the armature. 
For the ideal railgun $\Delta E=\Delta E_{mag} +\Delta E_{kin}$, with
both components being of the same magnitude. The energy that is stored in a PPS unit will be 
exhausted after supplying the driving electric power for a certain time. Assuming that the
amount of energy being stored in a PPS unit is just $\Delta E$ it follows that 
\begin{equation}
\Delta E = P \cdot \Delta t
\end{equation}
During a launch the required power increases, therefore the above power $P$ is the average power
during the short time interval $\Delta t$. The same holds for the velocity, within the
time $\Delta t$, the velocity increases, but an average velocity $v$ can be used to replace
the time period by
\begin{equation}
\Delta t= \frac{\Delta x}{v}
\end{equation}
With this equation one gets for the energy
\begin{equation}
\Delta E = P \cdot \frac{\Delta x}{v}
\end{equation}
As power is force times velocity, the equation can be rewritten as
\begin{equation}
\frac{\Delta E}{\Delta x} = F 
\end{equation}
This last equation can be interpreted in this way: If the force $F$ is constant, the 
railgun will consume the same amount of energy per barrel distance increment, 
regardless of the armature velocity. As it is one of the design goals to achieve a constant force 
and therefore acceleration of the armature, the PPS units trigger points have to be spaced by 
a constant distance. This "step-size" has to be adapted to the energy content of a PPS
unit. This insight is used as a guiding principle for the simulations being presented
in this paper.
\section{System Resistance}
\label{sys_res}
\begin{figure}[tb!]
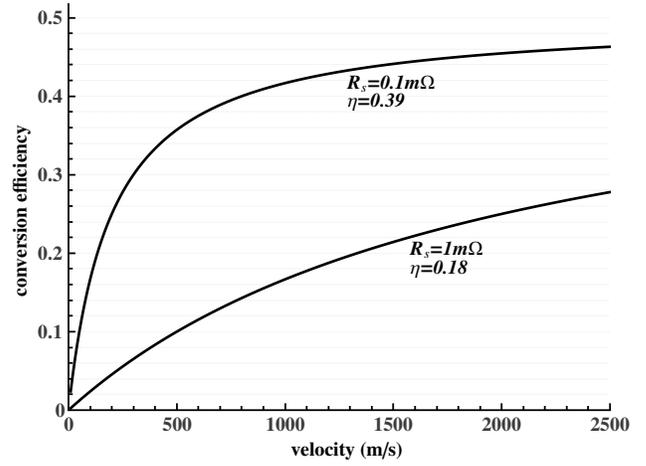

\centering
\resizebox{0.95\linewidth}{!}{

}
\caption{Power conversion efficiency for a L$^\prime$=0.5\,$\mu$H/m railgun at different system resistances.}
\label{conv_effi}
\end{figure}

In a railgun weapon system, all energy that is lost in the resistive part of the system can not be used for 
acceleration of the launch package. Therefore the system resistance has to be reduced as
much as possible. Especially for a multi-shot system the joule effect generates heat loads
which need to be removed from the weapon system at the cost of additional required energy,
thus further reducing the overall efficiency. The energy lost is the power spent during
the acceleration time. The following line of thought follows a discussion in \cite{loeffler}.
The power $P_{joule}$ that is converted into heat in the system due to the
system resistance $R_s$ is
\begin{equation}
P_{joule}=R_s \cdot i^2
\end{equation}
In this equation, the system resistance includes all resistances (from capacitors, coils, switches, cables,
connectors, rail resistance,{\ldots} ) and changes during the acceleration period. At the
same time the power that is required to sustain the kinetic acceleration process is force
times velocity, or
\begin{equation}
P_{kin}=\frac{1}{2}~L' i^2 v
\end{equation}
The same amount of power is required for the build-up of the magnetic field, thus the power related to the
acceleration can be written as
\begin{equation}
P_{acc}=L'i^2v
\end{equation}
The required power needed by the railgun during launch is $P=P_{acc}+P_{joule}$ and leads
to the relation that the fraction of the usable power for acceleration (conversion
efficiency) is
\begin{equation}
\frac{P_{acc}}{P}=\frac{P_{acc}}{P_{acc}+P_{joule}}=\frac{1}{1+\frac{R_s}{L'v}}
\label{conversion}
\end{equation}
If one is interested in the fraction of power that is converted into kinetic energy, the
formula \ref{conversion} is to be rewritten as
\begin{equation}
\frac{P_{kin}}{P}=\frac{P_{kin}}{P_{acc}+P_{joule}}=\frac{1}{2+\frac{2\cdot R_s}{L'v}}
\label{conversion_kin}
\end{equation}
This last two equations show the interesting behavior of a railgun: The efficiency is a
function of the velocity and increases with the velocity. They also show, that for a given railgun, 
it is of utmost importance to reduce the resistance as much as possible. This is of course stating 
the obvious: For a machine that consumes megaamperes the resistive losses become prohibitive if not the 
resistance is minimized.
In figure \ref{conv_effi} the equation \ref{conversion_kin} is evaluated for two different
constant system resistances and an
inductance gradient of 0.5\,$\mu$H/m. For a launch in the long range artillery scenario, the armature
is accelerated from stand-still to a velocity of 2500\,m/s. The implicit assumption is a
constant amplitude current pulse and no additional losses due to friction. For every velocity the
efficiency to convert the electrical power of the PPS into kinetic energy of the
projectile is shown in this figure for a system resistance of 0.1\,m$\Omega$ and 1\,m$\Omega$. 
This efficiency is a strong function of the velocity and the system
resistance. In the case of a DC current pulse, the acceleration is constant and the launch
efficiency is the arithmetic average of each of these curves. These efficiencies are
denoted in the figure by $\eta$. For a launcher with the above mentioned inductance gradient a 
system resistance of 1\,m$\Omega$ results in an efficiency of 18\% and at 0.1\,m$\Omega$
it reaches 39\%. It has to be mentioned that the system resistances used here are not the
same as the resistances in the simulation described later in this paper. There the
mentioned resistance is the resistance of one rack including the connection to the
railgun. As several racks are feeding the railgun in parallel, the overall resistance is
smaller than the used rack resistance.  Real world efficiencies (and also those in the
simulation described later) will deviate from these
calculated ones for several reasons: The
performance when using a real railgun degrades as mechanical friction and eddy currents occur,
but the magnetic energy stored in the inductance of the rails can be converted (at least
partly) into kinetic energy before shot-out by a drop in current amplitude, thus
increasing the efficiency. For the later effect, the
barrel needs to extend beyond the length at which all PPS units are exhausted.  
\section{Capacitor based PPS}
\begin{figure}[htb]
\centering
\includegraphics[width=.4\textwidth]{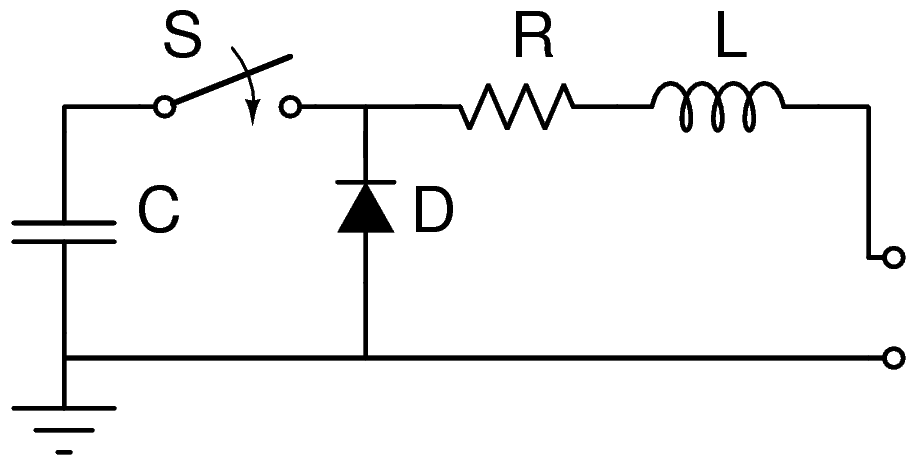}
\caption{\label{cap_rack}Schematic representation of a capacitor based PPS-rack.}
\end{figure}

A capacitor based PPS unit (here called rack) is composed out of a capacitor, a switch, a
pulse-forming coil, a crow-bar diode and a resistance.  A slightly simplified electrical
circuit diagram of 
such a rack is shown in figure \ref{cap_rack}. In this investigation,
the inductance of the pulse-forming coil contains all the inductances of the circuit
including the cables connecting the rack to the railgun. The same holds for the
resistance. The minor simplification of this circuit diagram is that it neglects the
slight change in total circuit inductance and resistance when the capacitor is exhausted
and the crow-bar diode becomes active. For practical purposes this small effect can be
ignored without effect on the resulting pulse height and shape. The peak
amplitude that is delivered by such a rack is calculated by using the equation:
\begin{equation}
I_{peak}=U_0 \cdot \sqrt{\frac{C}{L}}
\label{f1}
\end{equation}
By a careful selection of the capacitance and inductance the peak current amplitude of one rack can
be adjusted. For an individual rack, not only the current amplitude, but also the
rise time of the current pulse is of importance. This value is calculated by
\begin{equation}
t_{peak} = \frac{\pi}{2}\cdot\sqrt{LC} 
\label{f2}
\end{equation} 
and dependent on the two parameters inductance and capacitance, as well. As explained in
the previous section, to be able to
allow for a close to constant acceleration of the launch package, it is not sufficient to
use only one rack as PPS. Instead the current output of several racks are superimposed by
triggering the current release of the individual racks in a sequence governed by the
passage of the armature through the barrel.
\section{Electric Circuit Simulation with Capacitor based PPS}
\label{cap_sim}
\begin{figure}[tb!]
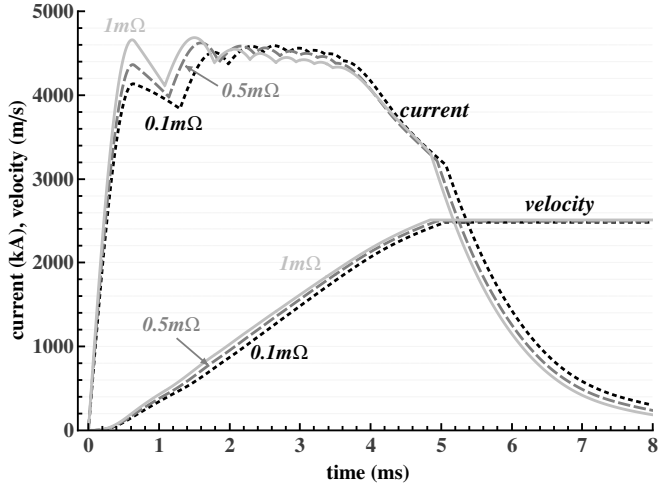

\centering
\resizebox{\linewidth}{!}{
%
}
\caption{Simulation results for the capacitor based racks.}
\label{cap_result}
\end{figure}

The simulation code using the NGSPICE \cite{ngspice} program which is routinely used to
evaluate the railguns at ISL was modified to simulate the 
proposed simple breech feed railgun with a 6.4\,m long barrel. The
inductance gradient L$^\prime$ is an input and was set to 0.5\,$\mu$H/m. For the rails,
copper material with a resistivity of 17\,n$\Omega$m was used. The resistance
gradient R$^\prime$ is calculated using the resistivity and the cross-section of the
rails. To account for the higher resistance due to the skin effect a factor
of 2 was used to increase the calculated rail resistance. With a width of
90\,mm and a rail height of 60\,mm and including the factor for the skin
effect a value of R$^\prime$ of 6.3\,$\mu\Omega$/m is used. This value evaluates for an average
resistance contribution from the rails of approx. 0.04\,m$\Omega$ for the 6.4\,m long
barrel. Friction was
taken into account by reducing the acceleration force by 10\%. Except for the inductance gradient, all assumptions 
of the relevant parameters are chosen
to be conservative. The starting position of the armature is located three times the caliber (0.27\,m) down
the barrel. This distance to the breech ensures that the armature experiences the full strength of the 
magnetic field. The total energy is distributed across
10 racks, which are individually connected to the breech of the railgun. Three racks are triggered to 
ramp up the current at launch start, the remaining 7 racks are triggered subsequently with the
passage of the armature through the barrel. To investigate the strong dependence of the
launch performance on the resistance as discussed in section \ref{sys_res}, three
cases are considered. The system performance was simulated using the value of the
resistance R in figure \ref{cap_rack} of 0.1\,m$\Omega$, 0.5\,m$\Omega$ and 1\,m$\Omega$. For each resistance the
capacitance and inductance of the rack and the trigger positions are modified to allow for an approximate equivalent
acceleration to the final velocity of 2500\,m/s (25\,MJ muzzle energy). The values for the resistance and
inductance as used in the simulation are listed in table \ref{tab_sim}. The different positions of the
armature to trigger the corresponding rack are listed in table \ref{tab_trig}. These
positions are determined to ensure a mostly flat current pulse shape. The results for the
three simulated cases are shown in figure \ref{cap_result}. A plateau of about 4.5\,MA of
current during most of the acceleration time is needed for the acceleration of the
armature. The dynamical nature of the circuit (increasing inductance and resistance)
during the launch makes an exact replication of the current pulse shape difficult for each
of the three investigated cases.
Therefore the optimization was stopped as soon as the final goal of 25\,MJ of muzzle energy was reached for
each case and the current peak amplitude did not exceed 4.5\,MA
for too long of a time. Due to the increasing resistance the energy stored increased from 53\,MJ (0.1\,m$\Omega$)
to 63\,MJ (0.5\,m$\Omega$) and finally to 
75\,MJ (1\,m$\Omega$). This translates to a launch efficiency of 47\%, 40\% and
33\%, respectively. The efficiencies are larger than the values in figure \ref{conv_effi},
as firstly, several racks do contribute to the current at the same time. During this
period, the rack resistances R are parallel, resulting in a smaller system resistance.
Secondly, the current pulse is not a DC pulse, instead the falling amplitude at the end of
the acceleration process leads to a conversion of the rail magnetic field into kinetic
energy.
\begin{table}
\begin{tabular}[tbh]{|l|*{6}{c|}}
\hline
R & E$_{rack}$ &C & L & t$_{peak}$ & I$_{max}$ \\
\hline
0.1\,m$\Omega$ & 5.3\,MJ & 22\,mF & 4\,$\mu$H &  0.47\,ms & 1.63\,MA \\
\hline
 0.5\,m$\Omega$ & 6.3\,MJ & 26\,mF & 4.6\,$\mu$H & 0.54\,ms & 1.65\,MA \\
\hline
 1\,m$\Omega$ & 7.5\,MJ & 31\,mF & 5\,$\mu$H & 0.62\,ms & 1.73\,MA \\
\hline
\end{tabular}
\vspace{1ex}
\caption{Parameters for 1 of 10 racks in 3 different PPS configurations. The capacitors
 are 22\,kV type.}
\label{tab_sim}
\end{table}
 
 \begin{table}
 \begin{tabular}[tbh]{|l|*{5}{r|}}
 \hline
 R $\mid$ Rack & 1 & 2 & 3 & 4 & 5\\
 \hline
 0.1\,m$\Omega$& 0.27\,m & 0.27\,m & 0.27\,m & 0.72\,m & 1.14\,m\\
 0.5\,m$\Omega$& 0.27\,m & 0.27\,m & 0.27\,m & 0.52\,m & 0.99\,m\\
 1\,m$\Omega$& 0.27\,m & 0.27\,m & 0.27\,m & 0.47\,m & 0.95\,m\\
 \hline
 \hline
 R $\mid$ Rack & 6 & 7 & 8 & 9 & 10\\
 \hline
 0.1\,m$\Omega$& 1.61\,m & 2.03\,m & 2.4\,m & 2.77\,m & 3.2\,m\\
 0.5\,m$\Omega$& 1.51\,m & 1.98\,m & 2.4\,m & 2.82\,m & 3.3\,m\\
 1\,m$\Omega$& 1.48\,m & 1.96\,m & 2.39\,m & 2.82\,m & 3.3\,m\\
  \hline
\end{tabular}
\vspace{1ex}
\caption{Positions to trigger the corresponding capacitor rack during the passage of the
 armature through the barrel.}
\label{tab_trig}
\end{table}
\section{Inductor Based PPS}
\label{coil_sim_section}
\begin{figure}[htb]
\centering
\includegraphics[width=.4\textwidth]{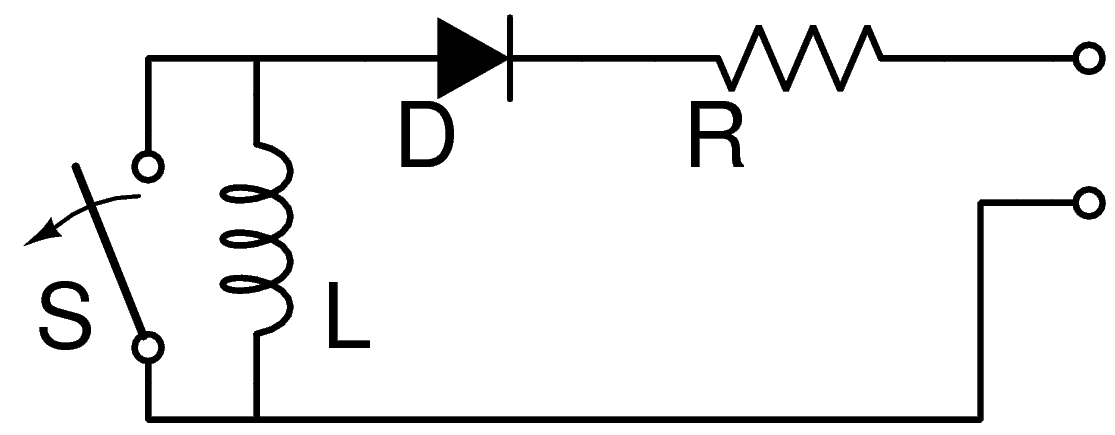}
\caption{\label{coil_rack}Schematic representation of an inductor based PPS-rack.}
\end{figure}

In a capacitor based PPS as shown in figure \ref{cap_rack}, the energy stored in a
capacitor is transferred to the pulse forming coil L and finally to the railgun. In an
inductively driven PPS, the coil does directly feed the railgun. In a coil
the energy content is
\begin{equation}
E_{ind}=\frac{1}{2} \cdot L I^2
\label{eqn_coil}
\end{equation}
When using the values for the inductance L and the current I$_{max}$ from
the table \ref{tab_sim} and the equation \ref{eqn_coil} one realizes that these compute
just to the energy content of a rack (5.3\,MJ, 6.3\,MJ and 7.5\,MJ). Therefore a
straight forward replacement of the capacitor based rack is to rearrange the circuit according 
to the schematic from figure \ref{coil_rack} and use the pulse-forming coil of the
capacitor based rack as storage inductance. This allows a one-to-one 
comparison of a railgun driven by rather similar
capacitor or inductor based PPS racks. Actually, when one does not want to change the
number of racks, there are no degrees of freedom available to deviate much from the
inductance value. The current amplitude is determined by the required acceleration of the projectile and
bound by the current carrying limit of the rails, and the total required energy by the
efficiency of the launch, which is dominated by the system resistance. 
For the racks from figure \ref{coil_rack} it is assumed that the switch S is closed and the coil L 
is charged with the corresponding current at the beginning of the launch process. The
charging system and its efficiency is not considered in this investigation. The
current is stored loss-free in the coil and the release of the energy is triggered by the
opening of the switch S. The discharge into the railgun is via the diode D and the resistance R. 
The purpose of the diode D is to disallow current from other racks
to enter the circuit. This prevents a recharging of a discharged coil from racks that are
triggered later during the launch process. When triggering a rack, the charged inductance
is switched to the railgun, which represents a variable and growing inductance. This reduces
the amplitude of the current from the storage coil and explains the smaller increments in
current amplitude per switched rack as the acceleration time progresses. 
\subsection{Simulation Results for Inductor Based PPS}
\begin{figure}[htb]
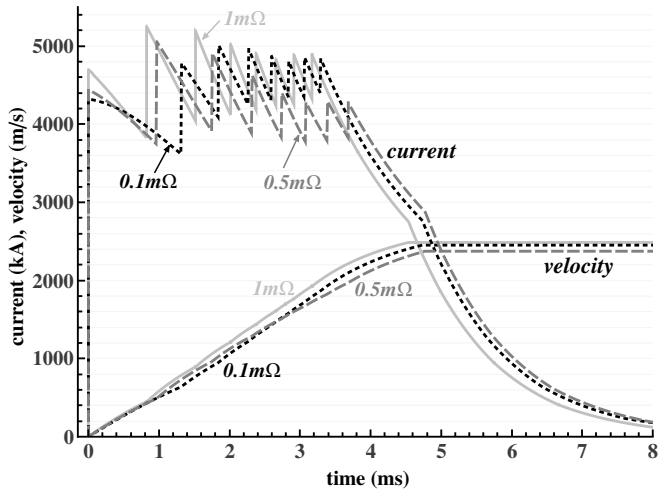

\centering
\resizebox{\linewidth}{!}{

}
\caption{\label{coil_sim}Simulation results using inductively driven racks.}
\end{figure}

For the simulation of the launcher performance, the simulation as described in section
\ref{cap_sim} was modified by replacing the capacitor based rack with the inductor based
rack in the SPICE circuit. After this, the cases for the system resistance of
0.1\,m$\Omega$, 0.5\,m$\Omega$ and 1\,m$\Omega$ were simulated. The results are shown in
figure \ref{coil_sim}. Inspecting the velocities of the projectile reveals, that the
reached velocities are of up to 5\% lower than the target
velocity of 2500\,m/s.
Inspecting the current trace, it becomes apparent, that the current amplitude variation is
not as smooth as in the simulation using capacitor based racks. The triggering of a charged
inductor equipped rack leads to a pronounced jump in the current amplitude. With the
increase in the railgun inductance during the launch process, the current variation
becomes smaller as the ratio total inductance to rack inductance becomes larger, but stays
still significant. In addition the maximum current amplitudes are above
5\,MA, higher than in the capacitor based simulations. The launch
efficiencies are 45\%, 36\% and 33\%. The relatively strong variation of the
current amplitude (peak-to-valley) of up to 1.4\,MA is a clear disadvantage for this
solution as this behavior translates to strong changes in the acceleration during the
launch and thus to a hard mechanical load for the launch package. As the current has to flow 
through the inductor-switch circuit at the time interval from charging of the coil up to the 
release of the rack energy, a real coil based PPS will suffer drastic losses during this time 
period. An effect that is neglected in this analysis, instead it is assumed that the
charging can take place infinitely quick just before the rack is triggered. 
\subsection{Increasing Rack Segmentation}
\begin{figure}[htb]
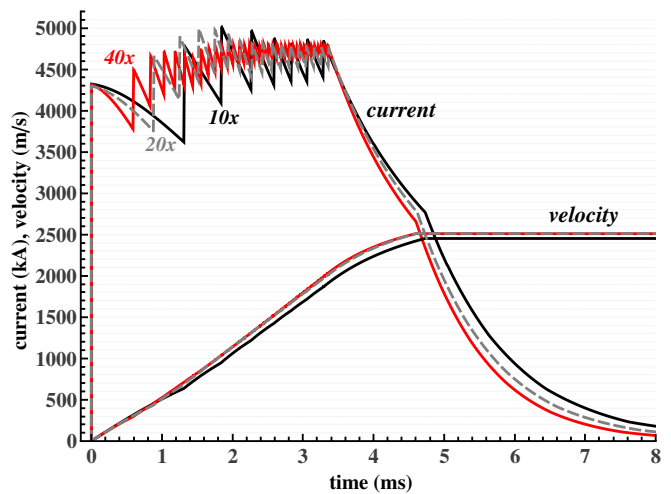

\centering
\resizebox{\linewidth}{!}{
%
}
\caption{\label{coil_sim_2}Simulation results for PPS subdivision into 10, 20 and 40 racks.}
\end{figure}

In the above simulation using inductor based racks with the same energy content as its
capacitor based counterpart it was shown that the current amplitude shows strong
variations during the launch. The sawtooth pattern results in an acceleration profile
that can be disadvantageous for delicate payloads (as for example a hypervelocity
projectile with moveable fins). One obvious idea to reduce the current variations is to increase
the number of racks by reducing the stored energy content per rack. To investigate the effect of smaller, 
but more energy portions
being fed to the railgun, the simulation was repeated using 20 and 40 racks. To first
order, one could think of simply varying the inductance value of the storage coil by a
factor corresponding to the ratio of the number of racks (i.e. using 2\,$\mu$H instead of
4\,$\mu$H when using 20 instead of 10 racks) and keeping the maximum current at the same
value. But as this simple scaling changes the ratio of the rack to railgun inductance,
the current amplitude in the railgun changes, too. Instead three parameters (inductance, current
amplitude and number of initially triggered racks) were varied to achieve a current amplitude of 
the same height for all three cases at the
starting time of the acceleration. The chosen parameters for the racks are shown in table
\ref{tab_coil_1}. As the inductance of the racks is reduced, it is required to increase
the number of racks that are triggered at acceleration start. As this number can not be
varied arbitrarily (racks can not be divided) the maximum current and the inductance were 
adopted using a trial-and-error method until the same initial current amplitude was
achieved in the three cases. The results of these simulations are shown in figure
\ref{coil_sim_2}. The velocity traces show the same behavior and within a small margin the 
end-velocity is the same for all the cases. Investigating the three current traces, one
can deduce that in fact the peak-to-valley amplitude of the sawtooth pattern becomes
smaller when the total energy is distributed to more individual racks. To get a
quantitative handle on this behavior, the arithmetic mean current value and the standard
deviation was calculated for every point in time of the simulation. The standard deviation
gives an information how far the data points vary from the mean value. Here we compare the
relative standard deviation of the current at the time when all racks have been fired (in the
figure \ref{coil_sim_2} the time when the current starts to drop rapidly at approx.
3.3\,ms). The values are 8\%, 5.5\% and 3.5\% for the simulation with 10, 20 and 40 racks,
respectively. This result clearly shows that it is possible to drastically reduce the
current (and acceleration) ripple in a coil based PPS driven railgun by using many small
racks.    
\begin{table}
\begin{tabular}[tbh]{|l|*{4}{c|}}
\hline
N$_{racks}$& E$_{rack}$ & L & I$_{max}$ & N$_{racks, initial}$\\
\hline
10 & 5.3\,MJ & 4\,$\mu$H & 1.63\,MA & 3\\
\hline
20 & 2.65\,MJ & 3.14\,$\mu$H & 1.3\,MA & 4\\
\hline
40 & 1.33\,MJ & 1.57\,$\mu$H & 1.3\,MA & 5\\
\hline
\end{tabular}
\vspace{1ex}
\caption{Individual rack parameters as implemented in the simulation using 10, 20 and 40 racks.}
\label{tab_coil_1}
\end{table}
\section{Energy Distribution}
\subsection{Capacitor Based PPS}
\begin{figure}[tb!]
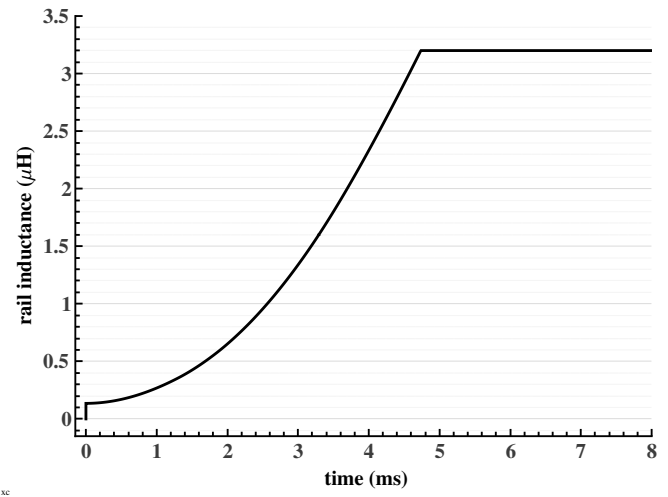

\centering
\resizebox{\linewidth}{!}{
xc

}
\caption{Rail inductance during launch.}
\label{rail_inductance}
\end{figure}

In the capacitor driven railgun system, the energy stored the capacitor is discharged into
the inductances of the system. These are the pulse-forming coil(s) of the PPS itself, the
inductance of the cables and the growing inductance of the launcher rails. The rail
inductance growth is shown in figure \ref{rail_inductance} for the launch with a coil
based PPS with 5.3\,MJ racks. As the acceleration is very similar in all simulated cases,
this figure is a good representation for the rail inductance. While the pulse forming
coils of one rack have an inductance of 4\,$\mu$H to 5\,$\mu$H, the rail inductance raises
up to a value of 3.2\,$\mu$H at muzzle exit. For the launch efficiency, it is of
importance to have as much as possible of the initially stored energy being converted into
kinetic energy of the projectile. In a breech fed railgun a non-negligible amount of
energy is being stored in the magnetic field in between the rails. In the DC-current case
this magnetic energy is of the same value as the kinetic energy of the projectile. In a
practical railgun the current has usually already started to decay and therefore the
rail magnetic field energy is at least partly converted into kinetic energy, thus
improving the launch efficiency. The inductances in the PPS itself transfer the stored
energy to the railgun. They are discharged by this energy transfer and by the system
resistance. In figure \ref{energy_ind} the distribution of the energy being stored in
these two inductances during the launch period are shown. Initially three racks are fired
simultaneously and  the energy is intermittently stored in the rack coils (in figure
\ref{energy_ind} the three traces showing the energy being stored in the coils of the
racks are marked by "E$_{ind,PPS}$", those for the rails by "E$_{ind,Rails}$"). 
As the armature progresses along the rails, the rail inductance 
becomes more important and more racks are fired. During this process the triggered racks are depleted from energy  
and more energy is stored in the rail magnetic field. The energy is being transferred from the 
PPS coils to kinetic energy of the projectile, magnetic energy of the rails and ohmic heat. 
The release points for the subsequently fired racks are close to equally spatially spaced. 
The explanation for the increase in the energy being stored in the active rack coils is
the following: As the projectile velocity increases, the projectile passes the distance in between two
subsequent trigger points faster than the charging/discharging time for a rack can
transfer the energy. After 4\,ms, the racks have all fired and the current through the railgun is driven
fully by the decaying magnetic field of the inductances. At shot-out there is approx.
5\,MJ of energy left in the PPS coils and between 16\,MJ to 18\,MJ still stored in the rail magnetic
field.
\subsection{Inductor Based PPS}
\begin{figure}[tb!]
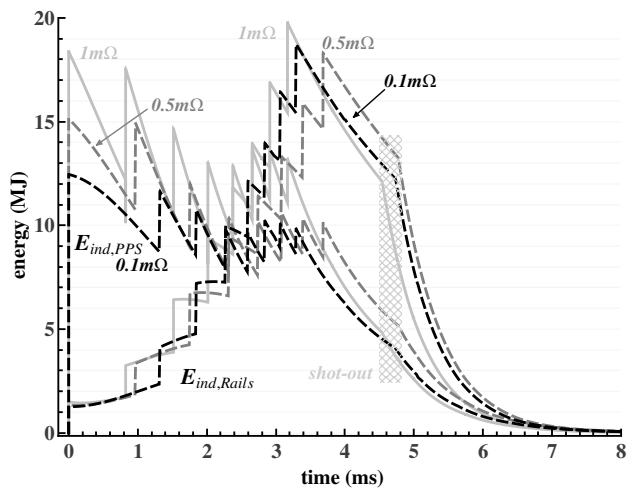

\centering
\resizebox{0.95\linewidth}{!}{
%
}
\caption{Coil based PPS: Energy stored in the inductances of the PPS and the rails for the three simulated
scenarios.}
\label{coil_energy_ind}
\end{figure}

In the inductor based PPS system, the discharge of the initially stored energy into the
railgun is more direct. At triggering the coil of the PPS rack is connected to the
railgun as resistive-inductive load. Discussing the discharge process shown in figure
\ref{coil_energy_ind} qualitatively, the
coil based system behaves overall as the capacitor based system. In the onset
of the acceleration process only very little energy is being transferred into the
inductance of the rails, but this changes rapidly with increasing launch time. The energy
being stored in the coils of the activated (triggered) racks is first dominating, but
becomes rapidly smaller once all racks had been triggered (after 3.2\,ms to 3.8\,ms
depending on the resistance). But the energy traces show a strong dependence on the
rack trigger process. Whenever a rack is triggered, the energy in the coils and rails changes 
rapidly and the traces develop a sawtooth like pattern. As such a behavior is not
beneficial for the launch (acceleration of payload, forces acting on barrel,{\ldots}) the
importance of a small resistance $R$ is again demonstrated by this figure. When
comparing the traces of the simulation for the 1\,m$\Omega$ and the 0.1\,m$\Omega$
resistances it can be seen that a smaller resistance and the resulting lower storage 
inductance results in a smoother energy transfer process. In this simulations the energy
being stored at shot-out in the PPS coils is in between 4\,MJ to 5\,MJ, while the rail
magnetic field stores in between 12\,MJ to 13\,MJ. The rail magnetic field energy at
shot-out is significantly smaller than in the capacity based system (by 4\,MJ to 5\,MJ)
and can be explained by the fact, that in the capacitor based system there is an
additional time to charge the pulse forming coils from the capacitor -- a delay that is
not existing in the inductor based system.
\begin{figure}[tb!]
\centering
\resizebox{0.95\linewidth}{!}{
%
}
\caption{Capacitor based PPS: Energy stored in the inductances of the PPS and the rails for the three simulated
scenarios.}
\label{energy_ind}
\end{figure}
\section{Summary and Conclusions}
Railguns are an attractive choice for long range deck guns. The combination of large
muzzle velocity with large muzzle energy allows to reach distances far in excess of
conventional artillery guns. The main problem of such a weapon system is
the lack of sufficiently small and lightweight electrical power supply capability on the platform. 
Therefore different
technologies are investigated as intermediate energy storage and power source. The most
mature solution are capacitor based PPSs. But capacitors have a large footprint and weight for a
given energy content. A PPS based on inductors might drastically improve the energy
density and is therefore an interesting alternative. In this investigation it was shown
that the electrical behavior of an inductor based PPS differs from the
capacitor based PPS. To first order, for a given PPS stored energy being split up into a number 
of identical racks, it is possible to accelerate a given projectile to the same
velocity using capacitor or inductor based racks. When exchanging charged capacitors for
charged coils, the efficiencies (kinetic energy divided by stored electrical energy in the
PPS) are approximately the same. The current amplitude trace is smoother for capacitor based racks
and therefore favorable compared to inductor based racks. For an inductor based PPS it is
possible to reduce the disadvantageous current ripple by increasing the 
 number of racks, by
making them smaller with respect to the stored energy. When taking into
account the charging of the rack and the hold time until discharge, capacitor based
systems seem to be favorable, at least in the configuration investigated here. Due to the
internal resistance of the storage inductor, inductor based racks can not be used to store
energy for a time longer than a fraction of the launch time without incurring prohibitive
losses.
\section*{Acknowledgment}
This research was supported by the French and German Ministries of Defense.
\end{document}